\documentclass[conference,compsoc]{IEEEtran}
\usepackage{ifpdf}
\ifCLASSOPTIONcompsoc
  \usepackage[nocompress]{cite}
\else
  \usepackage{cite}
\fi

\ifCLASSINFOpdf
  \usepackage[pdftex]{graphicx}
  \graphicspath{{../pdf/}{../jpeg/}}
  \DeclareGraphicsExtensions{.pdf,.jpeg,.png}
\else
  \usepackage[dvips]{graphicx}
  \graphicspath{{../eps/}}
  \DeclareGraphicsExtensions{.eps}
\fi
\usepackage{amsmath}
\usepackage{algorithmic}
\usepackage{array}
\ifCLASSOPTIONcompsoc
 \usepackage[caption=false,font=footnotesize,labelfont=sf,textfont=sf]{subfig}
\else
 \usepackage[caption=false,font=footnotesize]{subfig}
\fi
\usepackage{fixltx2e}
\usepackage{url}
\usepackage{enumitem}
\usepackage{hyperref}
\usepackage{multirow}
\usepackage{cite}
\usepackage{amsmath,amssymb,amsfonts}
\usepackage{algorithmic}
\usepackage{fancyhdr}
\usepackage{graphicx}
\usepackage{textcomp}
\usepackage{xcolor}
\def\BibTeX{{\rm B\kern-.05em{\sc i\kern-.025em b}\kern-.08em
    T\kern-.1667em\lower.7ex\hbox{E}\kern-.125emX}}

\begin{document}
%
\title{Su-RoBERTa: A Semi-supervised Approach to Predicting Suicide Risk through  Social Media using Base Language Models}

\author{
\IEEEauthorblockN{Chayan Tank}
\IEEEauthorblockA{MIDAS Lab\\ IIIT, Delhi\\ 
Email: chayan23030@iiitd.ac.in}
\and
\IEEEauthorblockN{Shaina Mehta}
\IEEEauthorblockA{MIDAS Lab\\ IIIT, Delhi\\ 
Email: shaina23139@iiitd.ac.in}
\and
\IEEEauthorblockN{Sarthak Pol}
\IEEEauthorblockA{MIDAS Lab\\ IIIT, Delhi\\ 
Email: sarthak23082@iiitd.ac.in}
\and 
\IEEEauthorblockN{Vinayak Katoch}
\IEEEauthorblockA{MIDAS Lab\\ IIIT, Delhi\\ 
Email: vinayak23105@iiitd.ac.in}
\and
\IEEEauthorblockN{Avinash Anand}
\IEEEauthorblockA{MIDAS Lab\\ IIIT, Delhi\\ 
Email: avinasha@iiitd.ac.in}
\and
\IEEEauthorblockN{Raj Jaiswal}
\IEEEauthorblockA{MIDAS Lab\\ IIIT, Delhi\\ 
Email: jaiswalp@iiitd.ac.in}
\and 
\IEEEauthorblockN{Rajiv Ratn Shah}
\IEEEauthorblockA{MIDAS Lab\\ IIIT, Delhi\\ 
Email: rajivratn@iiitd.ac.in}
}



%


\maketitle

\begin{abstract} In recent times, more and more people are posting about their mental states across various social media platforms. Leveraging this data, AI-based systems can be developed that help in assessing the mental health of individuals, such as suicide risk. This paper is a study done on suicidal risk assessments using Reddit data leveraging Base language models to identify patterns from social media posts. We have demonstrated that using smaller language models, i.e., less than 500M parameters, can also be effective in contrast to LLMs with greater than 500M parameters. We propose Su-RoBERTa, a fine-tuned RoBERTa on suicide risk prediction task that utilized both the labeled and unlabeled Reddit data and tackled class imbalance by data augmentation using GPT-2 model. Our Su-RoBERTa model attained a 69.84\% weighted F1 score during the Final evaluation. This paper demonstrates the effectiveness of Base language models for the analysis of the risk factors related to mental health with an efficient computation pipeline.

\end{abstract}

\begin{IEEEkeywords} Suicide Risk prediction, Large Language Models, RoBERTa, GPT-2, Reddit dataset, Data augmentation, Semi-supervised learning
\end{IEEEkeywords}

\IEEEpeerreviewmaketitle
\begin{figure*}[htbp]
    \centering
    \includegraphics[width=\textwidth]{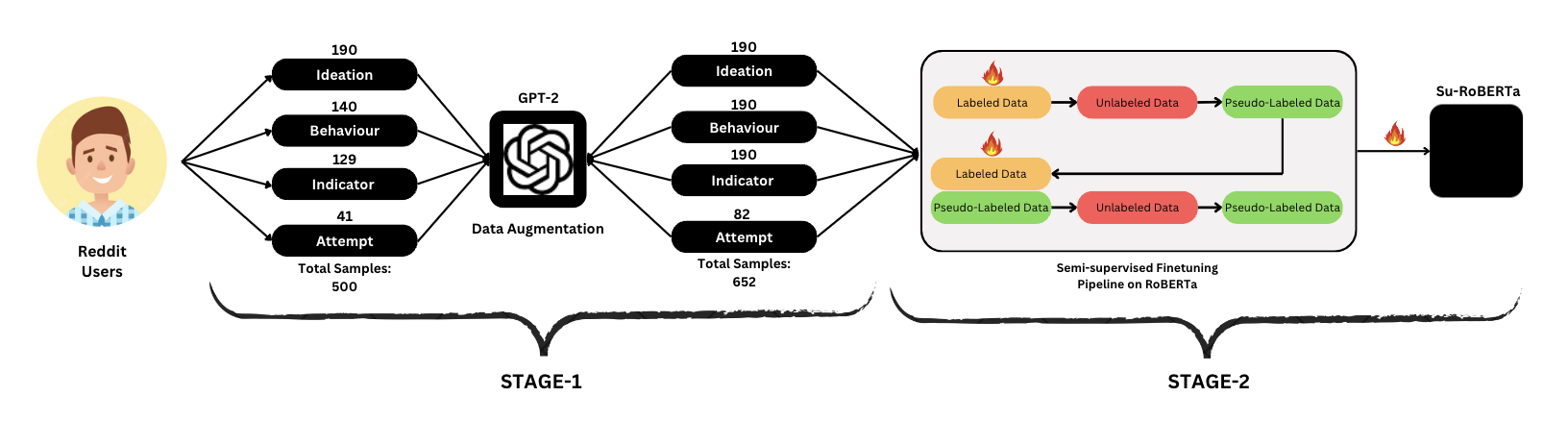}
    \caption{The Proposed Base Language Model approach for SuRoBERTa: Stage 1: Data augmentation of labeled samples using GPT-2 model. Stage 2: Proposed Semi-supervised Learning pipeline for Su-RoBERTa. The Fire symbol represents the model is being Fine-Tuned.}
    \label{fig:architecture_image}
\end{figure*}

\begin{figure*}[htbp]
    \centering
    \includegraphics[width=0.8\textwidth]{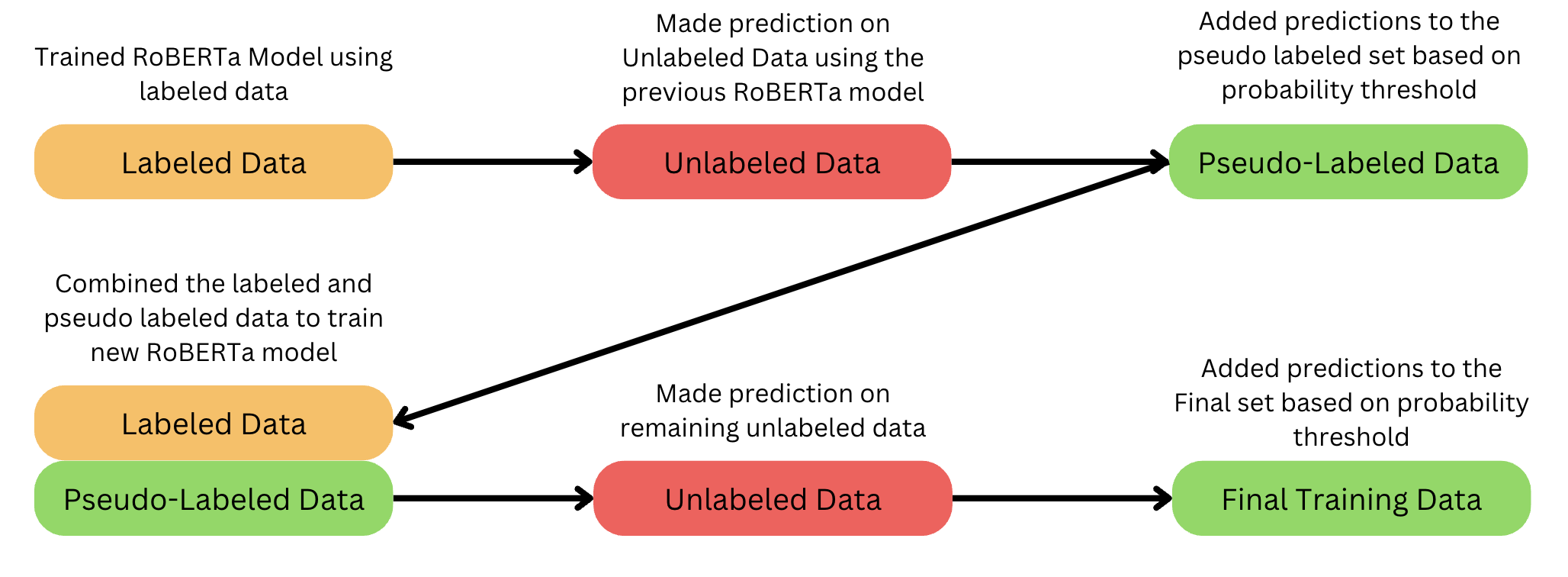}
    \caption{Su-RoBERTa Semi-supervised Fine-tuning pipeline mentioned in Stage 2 of Figure \ref{fig:architecture_image}}
    \label{fig:architecture_image_2}
\end{figure*}

\section{Introduction}




Suicide is one of the major public health problems with profound effects on the lives of individuals, families, and communities at large. The World Health Organization reports that an estimated 720,000 people die each year from suicide. It is one of the leading causes of death among young people and other vulnerable groups, such as refugees or migrants, as well as other sexual orientations\cite{b4}. Factors that result in suicidal behavior are multifaceted and complex. They include financial difficulties, interpersonal conflicts, chronic illness, and mental health disorders such as depression and anxiety. Knowledge of these contributing factors is critical in the development of effective prevention strategies. The traditional methods of suicide assessment rely on structured interviews and assessments, which consume a lot of time and are prone to personal biases.

Recently, a few studies show that people who are having suicidal thoughts use social media more frequently for sharing their mental state as well as suicidal thoughts  \cite{b5}. As a result, social media has become the major source for collecting data related to suicide for developing several solutions for suicide risk assessment. Using anonymity features in Social media Platforms such as Twitter and Reddit, users are encouraged to openly discuss their emotional states and mental health issues, which can provide rich insights for understanding suicidal ideation and facilitating timely interventions.

Over the years, many researchers have leveraged online data to model various aspects of mental health tasks, such as depression detection and suicidal ideation, by employing large language models (LLMs) for a deeper contextual understanding and reasoning \cite{b19} \cite{b20}. Through these researches, it is clear that LLMs show superior performance for text-based classification tasks in the mental health domain, but They are also computationally expensive to perform if fine-tuning is required for a specific task, our aim for this paper is to demonstrate that simpler language models with fewer parameters, being efficient can also provide a decent performance on same tasks, clearly not as well as bigger LLMs, but certainly more deployable on mobile compute or edge devices. With simplicity, efficiency, and deployability in mind, we propose Su-RoBERTa, our semi-supervised finetuned RoBERTa model on the suicide risk dataset revealed in the IEEE BigData 2024 Detection of Suicide Risk on Social Media Competition. To train this model, we also had to perform a compute-efficient way of data augmentation to tackle the data imbalance of this dataset. For this, we used the GPT-2-124M parameter model, leveraging its Superior generative capabilities for upsampling the sensitive suicide dataset while being computationally efficient.\\
The contribution of this study is threefold: 

\begin{itemize}
    \item\textbf{Data augmentation}: Demonstrated use of GPT-2 model for social media suicide data augmentation to mitigate class imbalance. 
    
    \item \textbf{Su-RoBERTa}: Proposing a fine-tuned RoBERTa model for suicide risk prediction task, along with a training pipeline using semi-supervised learning. 
    
    \item \textbf{Feasibility}: Demonstrated Effectiveness and efficiency of Base Language Models with fewer than 500M parameters over Large language models, typically exceeding 500M parameters for Mental health risk prediction tasks. 
\end{itemize}

\section{Literature Review}

Recently, several researchers have been working on automatic tools for the detection of suicide risk using machine learning, deep learning and base language models on the textual data obtained from social media websites. For instance, Matero et al. \cite{b43} leveraged the use of message, post ans use level statistics, BERT embeddings etc. and trained logistic regression and LSTM model with attention mechanism for suicide risk assessment. This model is also known as "Context-BERT" in several research work (such as in \cite{b42}) and achieved F1 score of 50 per cent on the dataset proposed by Shing et al. \cite{b23} during CLPsych 2019 competition \cite{b22}. Another approach is proposed by Allen et al. \cite{b21} leveraged the use of LIWC features for predicting the level of suicidal risk and trained their proposed CNN architecture using 10 Fold Cross Validation technique and achieved a macro average F1 score of 50 percent on test set of dataset proposed by Shing et al. \cite{b23} during CLPsych 2019 competition \cite{b22}. 

Bitew et al. \cite{b24} used weighted TD-IDF features, extracted emotion features from the text using the DeepMoji model and used them to train logistic regression and SVM model and combined all the features and models. They achieved the macro F1 score of 44.5 per cent on a test set of the suicide risk assessment dataset proposed by  Shing et al. \cite{b23} during the CLPsych 2019 competition \cite{b22} by using weighted TF-IDF vectors to train the logistic regression model. Priyamvada et al. \cite{b11} extracted textual embeddings from the posts of the Twitter dataset obtained from GitHub using Word2Vec and created a series of CNN-LSTM based architecture for suicide risk assessment and trained the model on it and achieved an accuracy score of 93.92 per cent. Mirtaheri et al., \cite{b6} proposed an AL-BTCN model consisting of LSTM, Bi-TCN (Bidirectional Temporal Convolutional Neural Network) and self-attention layer for suicide ideation detection among users. They extracted word embeddings from the BERT (Bidirectional Encoder Representations from Transformers) on Twitter and Reddit datasets obtained from GitHub and Kaggle and fed them into the AL-BTCN model for training. They achieved an F1 score of 92 per cent, 92 per cent and 95 per cent from two Twitter datasets and one Reddit dataset, respectively. Finally, Sawhney et al. \cite{b42} proposed a new approach for suicide risk assessment by leveraging the use of adversarial learning to train the deep learning architecture consisting post level embeddings obtained from BERT and performed modelling using LSTMs and Bi-LSTMs and achieved the F1 score of 64 per cent on a dataset created by Gaur et al. \cite{b9}. 

Interestingly, LLMs are also being applied in various other domains, such as the scientific and educational fields. In the scientific domain, LLMs have shown promise in tasks like citation generation \cite{b28} and grammatical error correction \cite{b30}, helping improve the accuracy and fluency of scientific writing. In the educational domain, their use spans across subjects like mathematics\cite{b34} and physics\cite{b29}, where researchers have developed methodologies to enhance learning experiences as well as Student engagement assessment \cite{b38}.

In the case of the suicide risk prediction task, there are also a few more types of research based on highlighting and summarizing the evidence of suicidal risk in social media posts, such as the works of Zhu et al. \cite{b25} who leveraged the use of Xiahai model, finetuned on ChatGLM3-6B model for evidence extraction task on suicide-related posts on Reddit. For better evidence extraction, they created their custom prompts and leveraged the use of Chain of Thoughts Prompting, One-Shot learning, text matching and regular expressions. From the extracted evidence, they created the summaries. They used their methodology on the Reddit dataset proposed by Shing et al. \cite{b23} and Zirikly et al. \cite{b22} achieving good performance in all the evaluation metrics proposed in the CLPsych 2024 competition \cite{b26}. Finally, Uluslu et al. \cite{b27} leverage the use of 4a -bit quantized Mistral 7B-Instruct-v0.2 model and perform prompt engineering in an iterative way to extract evidence and perform relevant string matching for extraction of correct evidence from suicidal posts of Reddit data proposed by Shing et al. \cite{b23} and Zirikly et al. \cite{b22}. They also performed summarization of evidence, they predicted the emotions of the posts using RoBERTa based emotion recognition model and summarization is performed using Prompt engineering and methodology similar to Retrieval Augmented Generation (RAG). They performed better in all the evaluation metrics proposed in the CLPsych 2024 competition \cite{b26}.

\section{Dataset Description}

The Dataset used for this challenge was proposed by Li et al. \cite{b1} and it consists of 2100 Reddit posts, out of which 2000 are reserved for model training and validation and 100 are reserved for testing purposes. The training set consists of 500 labeled Reddit posts and 1500 unlabeled Reddit posts, and the test set contains 100 posts whose labels are unknown to us as per the competition guidelines. The leaderboard position was determined by the evaluation of the revealed test set without labels, and the Final evaluation was done on an unrevealed test set.  

\begin{figure}[htbp]
    \centering
    \includegraphics[width=0.45\textwidth]{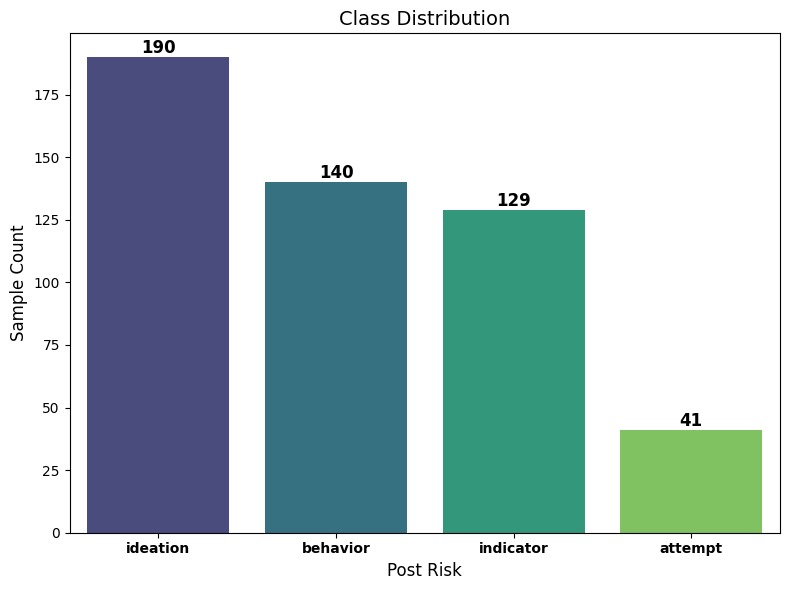}
    \caption{Original Sample distribution of the dataset}
    \label{fig:dataset_image}
\end{figure}

In the Labeled training samples, Each post is categorized into 4 categories which are: \\

\begin{itemize}
    \item \textbf{Ideation}
    These 190 posts contain suicide content, but there is no intention to die by suicide. They can speak about thoughts, feelings, or ideation for suicide, but they do not have the act or a more detailed suicide plan.

    \item \textbf{Behavior} 
    These 140 posts contain suicide content that speaks about suicide with a clear plan or intent to commit suicide. This category contains a more severe level of risk compared to ideation, as the user might have details on how they intend to act in this regard.
    
    \item \textbf{Indicator}
    These 129 posts do not contain contents that simply have suicidal intent. The language or tone may involve distress or general emotional issues, but nothing is apparent about mentioning suicide or any suicidal risk behavior.
    
    \item \textbf{Attempt}
    These 41 posts deal with the history of a past suicide attempt. The content may include stories of past suicide attempts made to commit suicide, meaning a high concern based on past activities.\\
\end{itemize}


\section{Data Augmentation}\label{aug}

\begin{figure*}[htbp]
    \centering
    \includegraphics[width=\textwidth]{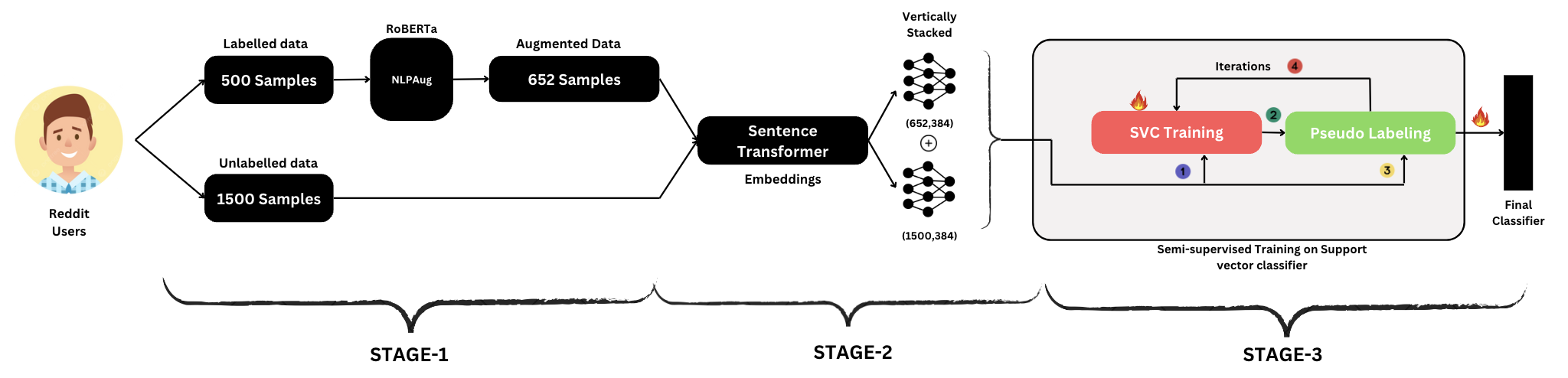}
    \caption{The Classical Semi-Supervised pipeline: Stage 1: Data augmentation of labeled samples using NLPAug + RoBERTa model. Stage 2: Using Sentence Transformer for extracting the embeddings of both Labeled and unlabeled samples and then Vertically stacking them for further processing. Stage 3: Classical Semi-supervised learning pipeline using SVMs. The Fire symbol represents the classifier being Trained.  }
    \label{fig:svc}
\end{figure*}

The class distribution in the dataset was imbalanced, as seen in Figure~\ref{fig:dataset_image}, with 'Attempt' being the minority class and 'Ideation' being the Majority class. The preliminary results were impacted by this imbalance. Hence, data augmentation was necessary for the dataset distribution and further modeling. We performed data augmentation using a RoBERTa-base and  GPT-2-base model to balance the dataset, which is mentioned in section \ref{4.2} and \ref{4.1} respectively and can also be seen as Stage 1 in Figures ~\ref{fig:architecture_image} and ~\ref{fig:svc}   respectively. 

\subsection{GPT-2}\label{4.2} 

\begin{figure}[htbp]
    \centering
    \includegraphics[width=0.45\textwidth]{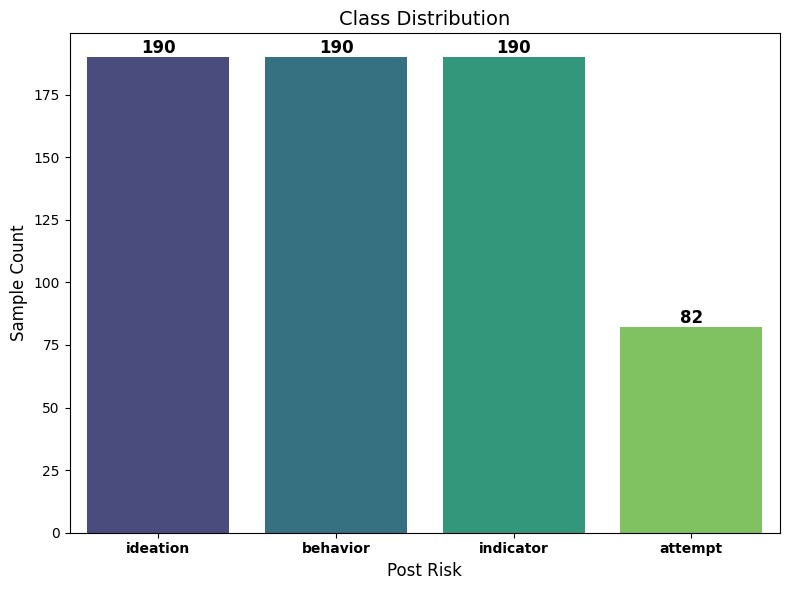}
    \caption{Augmented Sample Distribution of the dataset}
    \label{fig:augment_image}
\end{figure}

A GPT-2-Base 124M parameter language model was chosen, keeping computational efficiency and faster text generation in mind. As seen in Figure~\ref{fig:augment_image}, we up-sampled the minority classes 'Behavior' and 'Indicator' to 190 samples, equating them to 'Ideation', which was the majority class. The 'Attempt' class was upsampled to  82 samples as there were only 41 original samples of it; up-sampling it to 190 samples degraded the quality of the samples, so we augmented it to double the original size. The GPT-2 model was fine-tuned to generate synthetic posts for these Minority classes. This augmented dataset increased 152 samples using upsampling, and the final labeled training set contained 652 samples for training instead of 500 initially. This Tackled the class imbalance and increased the labeled training samples as well.


\subsection{RoBERTa}\label{4.1}
RoBERTa model proposed by Liu et al. \cite{b3}, an extension of BERT model mainly used for only Masked Language Modeling task and handles noisy datasets effectively. We performed the data augmentation by generating similar sentences from the minority classes using the RoBERTa base model, inspired by the approaches mentioned in \cite{b13} \cite{b14}, which is available in NLPAug \footnote{\url{https://github.com/makcedward/nlpaug}} library in Python. We have upsampled 'Behavior', 'Indicator' and 'Attempt' class to 190, 190 and 82 samples respectively, leading to final labeled training set contained 652 samples for training instead of 500 initially which is shown in Figure \ref{fig:augment_image}.



\section{Proposed Methodology}\label{methods}

The Objective of the competition was to classify the 100 test set samples into one of these 4 classes based on the patterns learned by leveraging the 500 labeled and 1500 unlabeled samples from the training dataset.

For our preliminary experimentations, We initially tried a classical semi-supervised approach by using NLPAug and SVM classifiers for data augmentation and modeling, respectively. This approach, although highly compute efficient is not that powerful, and its performance on evaluation is not that impressive. This compelled us to shift towards language models, mainly base language models such as BERT and RoBERTa, which are still more efficient than LLMs such as Llama-8B, Mistral models, GPT-3.5, etc. The augmentation for this case has also been done by a GPT-2 language model, which is a better generative model than the base language models.

\subsection{Classical Semi-supervised approach: Support Vector classifier (SVC)}

Given the size of the data and the large number of unlabeled training data, initially, we opted to use a classical semi-supervised learning approach inspired by \cite{b15} \cite{b16}, keeping computational efficiency as the goal. For the pre-processing of data, we have replaced emojis, converted the text to lowercase, removed special characters, HTML tags, and accented tags, and expanded acronyms. 

To remove the data imbalance, we performed data augmentation by generating similar sentences from the minority classes as mentioned in section \ref{4.1}. Then, textual embeddings are extracted using Sentence BERT proposed by Reimers et al. \cite{b17} using Sentence Transformers library in Python \footnote{\url{https://sbert.net/}}, which captures the semantics of the sentences. The labeled and unlabeled training data is combined and fed into the SVM classifier, and the model is trained using a semi-supervised algorithm from Scikit-learn library in python \footnote{\url{https://scikit-learn.org/stable/}} whose implementation is based on \cite{b2}. This pipeline can be seen in Figure~\ref{fig:svc} in 3 Stages.

\subsection{Base Language Model approach: Su-RoBERTa}

Even while leveraging the capabilities of Language models, our primary goal was to demonstrate comparable results to Large Language models greater than 500M parameters while minimizing computational demands. By using Base language models, we aimed to design an efficient, lightweight classifier pipeline capable of delivering predictions with reduced resource requirements, making it suitable for deployment on low-compute devices like mobile phones.

As discussed in section \ref{4.2}, we first up-sampled the minority classes and then used the augmented data along with the original samples to Finetune the RoBERTa model, Keeping computational complexity in mind, in an iterative semi-supervised approach presented in Figure~\ref{fig:architecture_image}. The RoBERTa model is generally better at understanding deep contextual meaning, while GPT-2 excels at generating text.

Initially, the RoBERTa 355M parameter model is fine-tuned on 4 class classification task using the augmented dataset. After fine-tuning the model on the labeled training samples, it is made to predict the classes of unlabeled training samples, which were 1500 in the count. Using these predictions, we pseudo-labeled the training data by following a greater than '0.33' probability policy, i.e., we only pseudo-labeled the samples that were predicted as a certain class with confidence of more than 0.33 probability, this probability threshold was decided by empirical experimentations. This ensured that the low-confidence predictions were not included as noise to further training of the model. 

These new Pseudo-labeled samples, along with the original samples, are now used as a training corpus to fine-tune the RoBERTa models again from scratch for better generalization. This approach is once again repeated, making a total of 2 iterations. The final Su-RoBERTa model was trained on the output pseudo-labeled data along with the original data; the sample count for final training included the whole of 2000 samples in the dataset available for training. This Fine-tuning pipeline has been shown in Figure \ref{fig:suRoBERTa}. The finetuning of the Model was performed using AdamW optimizer, with a batch size of 8 samples running for 10 epochs on a 24 GB NVIDIA GeForce RTX 3090. The whole semi-supervised finetuning pipeline took 30 minutes to complete. 

This demonstrates the Low compute required to fine-tune such a model for mental health prediction tasks. In the future, such training pipelines can also be deployed on mobile devices that are not that compute-heavy to train or finetune LLMs. 

\section{Evaluation}

\begin{table}[ht]
\caption{\small Few Shot Prompt for Labeling the Test Set using GPT-4} 
\centering
\begin{tabular}{|>{\raggedright\arraybackslash}m{8cm}|}
\hline
\\\textbf{\fontsize{10}{14}\selectfont \textcolor{blue}{System Prompt:}} \\
\\\textbf{You are the psychologist and your task is to predict the suicidal risk level from the given posts among 4 categories that are ‘indicator’,  ‘ideation’, ‘behavior’ and ‘attempt’. The sample posts and their class is given below: } \\

\\\textbf{Post: \texttt{<post>}} 
\\\textbf{Label: \texttt{Ideation}} \\
\\\textbf{Post: \texttt{<post>}} 
\\\textbf{Label: \texttt{Behaviour}} \\
\\\textbf{Post: \texttt{<post>}} 
\\\textbf{Label: \texttt{Indicator}} \\
\\\textbf{Post: \texttt{<post>}} 
\\\textbf{Label: \texttt{Attempt}} \\
\\
\hline
\\\textbf{\fontsize{10}{14}\selectfont \textcolor{red}{User Prompt:}} \\
\\\textbf{Now I am giving you a new post and your task is to predict the suicidal risk level from the given posts among 4 categories that are ‘indicator’,  ‘ideation’, ‘behavior’ and ‘attempt’. While predicting, you take reference from the examples given above. Here is the post:} \\
\\\textbf{Post: \texttt{<enter the unlabeled post here>}} \\
\\
\hline
\\\textbf{\fontsize{10}{14}\selectfont \textcolor{teal}{Model Response:}} \\
\\\textbf{Label: \texttt{<label>}} \\\\
\hline
\end{tabular}
\label{table_2}
\end{table}

As we did not have access to the test set labels on which our final evaluation was done, For evaluating our methodologies and models, we have used the state-of-the-art (SOTA) proprietary Large language model GPT-4 and used it through the ChatGPT platform to manually label the 100 test samples by performing few-shot prompting using the training data. The Few shot prompting has been shown in \ref{table_2} where we have done 4 shot prompting by including one sample from every class from the training set and then prompted the model to generate a response for every input sample from the test set. Now we have treated this pseudo-labeled test set as our evaluation ground truth to benchmark our models against. It is clear that GPT-4 itself cannot be perfect for this specific task, but given its SOTA performance on zero and few shot tasks, it is logical to assume that it can be decent enough to be used for baseline comparison of our models and will help in comparison of showing the capabilities of Base language model in contrast to the SOTA Large language model performance. 

\begin{figure}[htbp]
    \centering
    \includegraphics[width=0.45\textwidth]{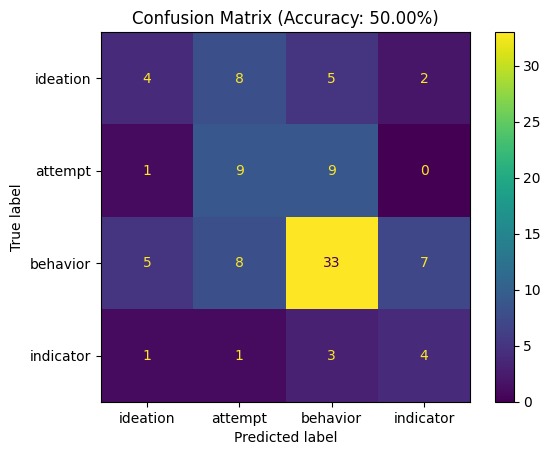}
    \caption{Su-RoBERTa evaluated against GPT-4 Labeled test set}
    \label{fig:suRoBERTa}
\end{figure}

\begin{figure}[htbp]
    \centering
    \includegraphics[width=0.45\textwidth]{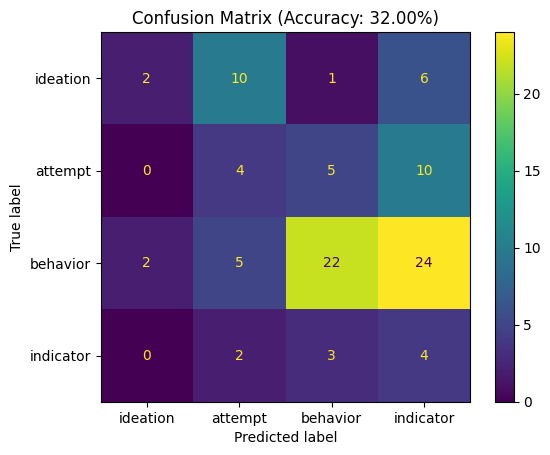}
    \caption{Support vector classifier evaluated against GPT-4 Labeled test set}
    \label{fig:svm}
\end{figure}

The evaluation of Su-RoBERTa against the GPT-4 labeled test set is shown in the confusion matrix \ref{fig:suRoBERTa}, and the evaluation of the Support vector classifier against it is given in \ref{fig:svm}. As it can be seen from confusion matrices, the classical SVM baseline model achieved an accuracy of 32\%, which is decent given that there are 4 classes and only 100 samples. Hence, 32 samples directly matched the GPT-4 labeled test set; It is also to be noted that the 'behavior' class has been mislabeled a lot to the 'Indicator' class.

Similarly, the accuracy of Su-RoBERTa against GPT-4 labels is 50\%, which implies that at least 50 samples directly match the GPT-4 labeled data, demonstrating that it is indeed better at predicting suicide risk when the base language model is used than the classical approach. This also shows that although SOTA LLM GPT4 has 50 unmatched predictions still the Base language model can be comparable to its performance, as it's clear that even GPT-4 labeled data might not completely be the exact class distribution as the original test set labels. Hence, this metric indicated Su-RoBERTa as a better-performing model.

\section{Results}

This section discusses the Final results achieved from both the approaches discussed in section \ref{methods}. Both the Su-RoBERTa model and the Support vector classifier are evaluated using the weighted F1 scores as per the competition guidelines. The results of both approaches are given in table \ref{table_1}. The test set of the competition was not revealed, and the evaluations have been done on the leaderboard of the challenge.

\begin{table}[ht]
    \centering
    \caption{Weighted F1 Scores after Preliminary Evaluation and Final Evaluation of Proposed Approaches on Test Set}
    \label{table_1}
    \renewcommand{\arraystretch}{1.2} 
    \setlength{\tabcolsep}{8pt} 
    \begin{tabular}{|l|c|c|}
        \hline
        \multirow{2}{*}{\textbf{Models}} & \multicolumn{2}{c|}{\textbf{Weighted F1 Score}} \\ \cline{2-3}
                                         & Preliminary Evaluation & Final Evaluation \\ \hline
        SVM                         & 50.52\%   & -  \\ \hline
        Su-RoBERTa                          & 61.31\%   & 69.84\%  \\ \hline
    \end{tabular}
\end{table}

Referring to the evaluation table \ref{table_1}, the SVM model achieved a Weighted F1 Score of 50.52 percent, whereas the Su-RoBERTa model achieved a Weighted F1 Score of 61.31 percent during preliminary evaluation. Based on the Weighted F1 Scores of both models, we have submitted the Su-RoBERTa model for final evaluation. On final evaluation, the Su-RoBERTa model achieved the Weighted F1 Score of 69.84 percent, which led us to the 10\textsuperscript{th} rank in the leaderboard of the competition.

\section{Conclusion}

The paper demonstrates the superiority of language models for social media suicidal risk prediction, leveraging smaller language models, i.e., base language models such as GPT-2 and RoBERTa, to be made into efficient language classifiers that are low-compute and deployable. These models achieved a decent performance on the challenge leaderboard, however, these still need to be improved for real-time mobile applications. In the future, we will focus on using techniques such as prompt engineering to enhance the LLMs' performance and use some advanced Large language models such as GPT-4o, OpenAI-o1 and Llama-3.2 models for data augmentation, labeling, and finetuning to improve the predictions on such mental health tasks. 

\section{Future Scope}

For a richer understanding, the inclusion of Multi-modal datasets for suicide and mental health analysis is crucial, Combining audio and visual data along with text for a complete Contextual overview. Social media platforms have evolved significantly in recent years. The shift towards multimedia content to platforms such as TikTok, Instagram, and YouTube Shorts popularizing the use of short video clips has provided new opportunities for analyzing social media data. To effectively assess the mental health of individuals based on these rich, multi-modal data forms, developing new architectures capable of understanding and processing multiple modalities is now essential. More metadata can be included in the analysis of mental health, such as sleep and stress detected through a wearable tracker and heart rate detection systems. 

Explainability analysis of these architectures and models is also crucial for mental health tasks as this type of data is sensitive, and the predictions need an underlying reason for the classification so that they can be verified by trained clinicians.






%

\end{document}